\documentclass[sigconf,screen]{acmart}
\AtBeginDocument{%
  \providecommand\BibTeX{{%
    \normalfont B\kern-0.5em{\scshape i\kern-0.25em b}\kern-0.8em\TeX}}}

\setcopyright{acmlicensed}
\acmPrice{15.00}
\acmDOI{10.1145/3540250.3549081}
\acmYear{2022}
\copyrightyear{2022}
\acmSubmissionID{fse22main-p37-p}
\acmISBN{978-1-4503-9413-0/22/11}
\acmConference[ESEC/FSE '22]{Proceedings of the 30th ACM Joint European Software Engineering Conference and Symposium on the Foundations of Software Engineering}{November 14--18, 2022}{Singapore, Singapore}
\acmBooktitle{Proceedings of the 30th ACM Joint European Software Engineering Conference and Symposium on the Foundations of Software Engineering (ESEC/FSE '22), November 14--18, 2022, Singapore, Singapore}

\usepackage{pifont}
\usepackage{xcolor}
\usepackage{threeparttable}
\usepackage{booktabs}
\usepackage{multirow}
\usepackage{relsize}
\usepackage{caption}
\usepackage{subcaption}
\usepackage{url}
\renewcommand{\paragraph}[1]{\vskip 0.05in \noindent {\bf #1}}

\begin{document}

%%
%% The "title" command has an optional parameter,
%% allowing the author to define a "short title" to be used in page headers.
% anonymous title for double blind review
% \title{CodeReviewer: Pre-Training for Automating Code Review Activities}
\title{Automating Code Review Activities by Large-Scale Pre-training}

%%
%% The "author" command and its associated commands are used to define
%% the authors and their affiliations.
%% Of note is the shared affiliation of the first two authors, and the
%% "authornote" and "authornotemark" commands
%% used to denote shared contribution to the research.
\author{Zhiyu Li}
\authornote{Equal contribution.}
\authornote{Work done during internship at Microsoft Research Asia.}
\affiliation{%
  \institution{Peking University}
  \country{China}
  }
\email{AkinoLi@pku.edu.cn}

\author{Shuai Lu}
\authornotemark[1]
\affiliation{%
  \institution{Microsoft Research Asia}
  \country{China}
}
\email{shuailu@microsoft.com}

\author{Daya Guo}
\authornotemark[2]
\affiliation{%
  \institution{Sun Yat-sen University}
  \country{China}
}
\email{guody5@mail2.sysu.edu.cn}

\author{Nan Duan}
\authornote{Corresponding author is Nan Duan.}
\affiliation{%
  \institution{Microsoft Research Asia}
%   \city{Beijing}
  \country{China}
}
\email{nanduan@microsoft.com}

\author{Shailesh Jannu}
\affiliation{
  \institution{LinkedIn}
  \country{USA}
}
\email{sjannu@linkedin.com}

\author{Grant Jenks}
\affiliation{
  \institution{LinkedIn}
  \country{USA}
}
\email{gjenks@linkedin.com}

\author{Deep Majumder}
\affiliation{
  \institution{LinkedIn}
  \country{USA}
}
\email{dmajumder@linkedin.com}

\author{Jared Green}
\affiliation{
  \institution{LinkedIn}
  \country{USA}
}
\email{jagreen@linkedin.com}

\author{Alexey Svyatkovskiy}
\affiliation{
  \institution{Microsoft DevDiv}
  \country{USA}
}
\email{alsvyatk@microsoft.com}

\author{Shengyu Fu}
\affiliation{
  \institution{Microsoft DevDiv}
  \country{USA}
}
\email{shengyfu@microsoft.com}

\author{Neel Sundaresan}
\affiliation{
  \institution{Microsoft DevDiv}
  \country{USA}
}
\email{neels@microsoft.com}

%%
%% By default, the full list of authors will be used in the page
%% headers. Often, this list is too long, and will overlap
%% other information printed in the page headers. This command allows
%% the author to define a more concise list
%% of authors' names for this purpose.
% \renewcommand{\shortauthors}{Zhiyu Li and Shuai Lu, et al.}
\renewcommand{\shortauthors}{Z.Li, S.Lu, D.Guo, N.Duan, S.Jannu, G.Jenks, D.Majumder, J.Green, A.Svyatkovskiy, S.Fu and N.Sundaresan}

%%
%% The abstract is a short summary of the work to be presented in the
%% article.
\begin{abstract}
Code review is an essential part to software development lifecycle since it aims at guaranteeing the quality of codes. Modern code review activities necessitate developers viewing, understanding and even running the programs to assess logic, functionality, latency, style and other factors. It turns out that developers have to spend far too much time reviewing the code of their peers. 
Accordingly, it is in significant demand to automate the code review process. 
In this research, we focus on utilizing pre-training techniques for the tasks in the code review scenario. We collect a large-scale dataset of real-world code changes and code reviews from open-source projects in nine of the most popular programming languages. 
% By proposing four pre-training tasks which are tailored specifically for better understanding of code diffs and reviews, we continue pre-train the Text-to-Text Transfer Transformers on Code (CodeT5) model.
To better understand code diffs and reviews, we propose CodeReviewer, a pre-trained model that utilizes four pre-training tasks tailored specifically for the code review scenario.
% To evaluate our model, we establish a high-quality benchmark dataset based on our collected data for code review activities, covering tasks such as code change quality estimation, which predicts whether a code change can be accepted, review comment auto-generation and code refinement, which is based on a code diff and its corresponding review. 
To evaluate our model, we focus on three key tasks related to code review activities, including code change quality estimation, review comment generation and code refinement.
Furthermore, we establish a high-quality benchmark dataset based on our collected data for these three tasks and conduct comprehensive experiments on it.
The experimental results demonstrate that our model outperforms the previous state-of-the-art pre-training approaches in all tasks. Further analysis show that our proposed pre-training tasks and the multilingual pre-training dataset benefit the model on the understanding of code changes and reviews.
\end{abstract}

%%
%% The code below is generated by the tool at http://dl.acm.org/ccs.cfm.
%% Please copy and paste the code instead of the example below.
%%


% \ccsdesc[500]{Computer systems organization~Embedded systems}
% \ccsdesc[300]{Computer systems organization~Redundancy}
% \ccsdesc{Computer systems organization~Robotics}
% \ccsdesc[100]{Networks~Network reliability}

%%
%% Keywords. The author(s) should pick words that accurately describe
%% the work being presented. Separate the keywords with commas.

%% A "teaser" image appears between the author and affiliation
%% information and the body of the document, and typically spans the
%% page.
% \begin{teaserfigure}
%   \includegraphics[width=\textwidth]{sampleteaser}
%   \caption{Seattle Mariners at Spring Training, 2010.}
%   \Description{Enjoying the baseball game from the third-base
%   seats. Ichiro Suzuki preparing to bat.}
%   \label{fig:teaser}
% \end{teaserfigure}

%%
%% This command processes the author and affiliation and title
%% information and builds the first part of the formatted document.
\maketitle

\section{Introduction}
Code review, a process of manually inspecting source code by teammates other than the code author, is recognized as a critical part of the software development lifecycle \cite{fagan2002inspect}. 
% Figure \ref{fig:code_review} depicts a general process of code review. When a pull request to a prct is created, it will trigger the code review process. Reviewers should fully inspect the code to check whether there exist issues with it. If all the requirements are satisfied, the pull request would be accepted and merged into the project. Otherwise, reviewers would raise comments to the code author. After updating the codes, the author can request another review. 
% Since Fazen \cite{fagan2002design} formalized a highly structured process for code reviewing -- code inspection, which is formal but cumbersome, 
Studies have shown the huge benefits of code reviews \cite{sadowski2018modern,bacchelli2013expectations,ackerman1984software,rigby2013convergent}. Compared with the traditional code review process formalized by Fagan in 1976 \cite{fagan2002design}, which avoids introducing errors and defects but is cumbersome, modern code review activities involve fewer formal requirements and aim at fully understanding code changes \cite{beller2014modern}. Given its benefits, code review has been widely adopted in both open-source and industrial projects. As shown by \citet{xin2016mining}, in open-source projects, such as Qt, there are ten thousand reviews taking place every month (\textasciitilde22k reviews in case of Qt). However, it's not free to take all advantages of code reviews. To make a correct judgement whether to accept the pull request, developers have to put much time and effort into code reviewing, checking code from every aspect, including logic, functionality, complexity, code style, documentation, etc. For example, \citet{xin2016mining} report that only 1,437 reviewers have given more than 1M reviews in 4 years in project Qt. Accordingly, it is in significant demand to automate the code review process.

% \begin{figure}
%     \centering
%     \includegraphics[width=\linewidth]{figs/code_review.pdf}
%     \caption{A general process of code review.}
%     \label{fig:code_review}
% \end{figure}

\begin{figure*}
    \centering
    \includegraphics[width=0.95\textwidth]{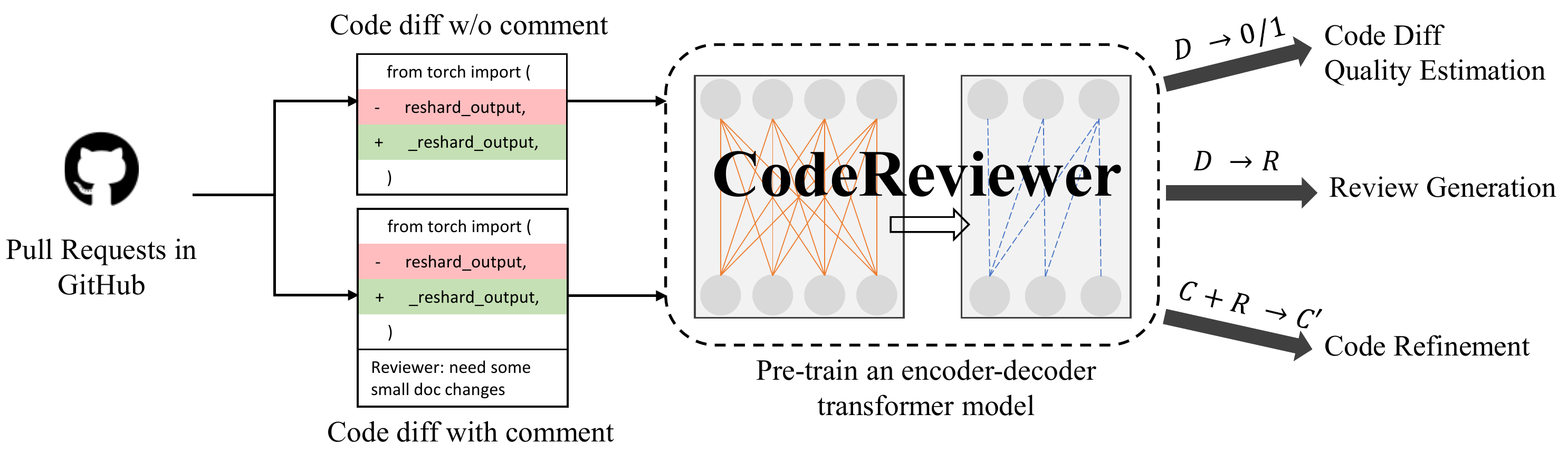}
    \caption{Overview of the workflow of CodeReviewer.}
    \label{fig:overview}
\end{figure*}

Many researchers have explored ways to assist reviewers and committers (i.e., code authors) to reduce their workload in the code review process, such as recommending the best reviewer \cite{thongtanunam2015should,chouchen2021whoreview}, recommending or generating the possible review comments \cite{gupta2018intelligent,tufano2021towards,tufano2022using} and even revising the code before submitting it for review \cite{tufano2022using}. This paper shares the same goal to automate some specific tasks related to code review. We target three scenarios from both reviewers' and committers' perspectives. The first is called \textit{code change quality estimation}, aiming to predict whether a code diff needs a review comment. It assists reviewers to pick a code diff chunk that might have issues among a large number of code chunks. The second task is \textit{review generation} which can dramatically reduce the time cost for the reviewer. The last scenario is for committers, which is \textit{code refinement} according to the previous  code and reviewer's comment.

Motivated by the wide adaptation of deep learning (DL) techniques for software engineering and the rapid development of pre-training techniques, we leverage pre-training for automating code review activities.
Recently, researchers have proposed many pre-trained models on source code \cite{feng2020codebert,guo2021graphcodebert,wang2021codet5,chen2021evaluating}. However, they can hardly handle the code review process. \citet{chen2021evaluating} propose Codex, a large scale pre-trained model based on GPT-3 \cite{brown2020language}. Codex has been proven to be the most powerful model for code generation tasks. Since it is trained on the source code files in raw format, it knows very little about code review. We demonstrate that it cannot generate any meaningful comment in the \textit{review generation} task based on our evaluation.
\citet{tufano2022using} attempt to use a pre-trained model for code review automation. However, their pre-training dataset is collected from Stack Overflow and CodeSearchNet \cite{husain2019codesearchnet}, which is not directly related to code review process. 
% Furthermore, they also take normal source code as model input without utilizing the special format of diff file to understand code changes better.
Furthermore, they take the normal form of source code as the model input, ignoring the special format of code diff that can help the model better understand code changes.

To tackle the problems, we pre-train \textbf{CodeReviewer}, an encoder-decoder transformer model. Different from \citet{tufano2022using}'s work,
CodeReviewer is pre-trained on a large dataset in code review scenario, consisting of code diff hunks and code review comments. We propose four pre-training tasks, including diff tag prediction, denoising code diff, denoising review comment, and review comment generation to make CodeReviewer better understand code diffs and generate review comments. Figure \ref{fig:overview} shows the overview of the workflow of our CodeReviewer.

% To obtain a pre-trained model that is suitable for the code review process, 
To make CodeReviewer more suitable for the code review process,
we build a large-scale dataset of code changes and corresponding review comments. 
% The importance of data in DL research, particularly pre-training research, has long been recognized. 
Such code review data is collected from GitHub pull requests in high-quality projects with high stars covering nine of the most popular programming languages. 
Using GitHub code review data, we create pre-training and benchmark datasets for three tasks related to the code review process. 
To the best of our knowledge, it is the largest multilingual code review dataset with complete information of code changes and reviews.
% This dataset will be open-sourced in order to facilitate the development of code review automation.

% Pre-trained on the above dataset, we obtain an encoder-decoder transformer model -- \textbf{CodeReviewer}. Different from \citet{tufano2022using}'s work,
% CodeReviewer is pre-trained on a large dataset in code review scenario, consisting code diff chunks and code review comments. We propose four pre-training tasks (diff tag prediction, masked line generation, review comment denoising and review generation) to make CodeReviewer better understand code diffs and generate review comments. Figure \ref{fig:overview} show the overview of the workflow of our CodeReviewer.

We further evaluate our CodeReviewer model on our benchmark dataset. Compare with previous state-of-the-art (SOTA) generative models for code, experimental results show that our model outperforms previous works on all three tasks. Further analysis proves the effectiveness of our proposed pre-training tasks and the multilingual high-quality dataset.

To summarize, the contributions of this work are:
\begin{itemize}
    \item The first pre-trained model that takes code diffs as input in the code review scenario. 
    \item Novel pre-training tasks for better code changes understanding and generation.
    % \item A pre-trained model for automating three different code review tasks. It is the first pre-trained model that takes code diffs as input in the code review scenario.
    % \item Novel pre-training tasks for better code changes understanding and generation.
    \item A large-scale code review related pre-training dataset and a high-quality benchmark dataset for evaluation in nine programming languages.
    \item A comprehensive evaluation of CodeReviewer and previous SOTA models. 
    % Our model has broad applicability in different programming languages. 
    Our dataset, code, and model are released
    % \item Our dataset, code, and model are open-sourced to facilitate the development of code review automation
    % % \footnote{https://doi.org/10.5281/zenodo.6355569}.
    \footnote{\url{https://github.com/microsoft/CodeBERT/tree/master/CodeReviewer}}.
\end{itemize}

\section{Code Review Automation Tasks}
In this section, we give a formulated description of the code review process, provide the formal definitions of three key tasks abstracted from the code review process, and clarify the data format of the tasks.
The employed symbols are listed in Table \ref{tab:notation}. 

\begin{figure*}
    \centering
    \includegraphics[width=0.95\textwidth]{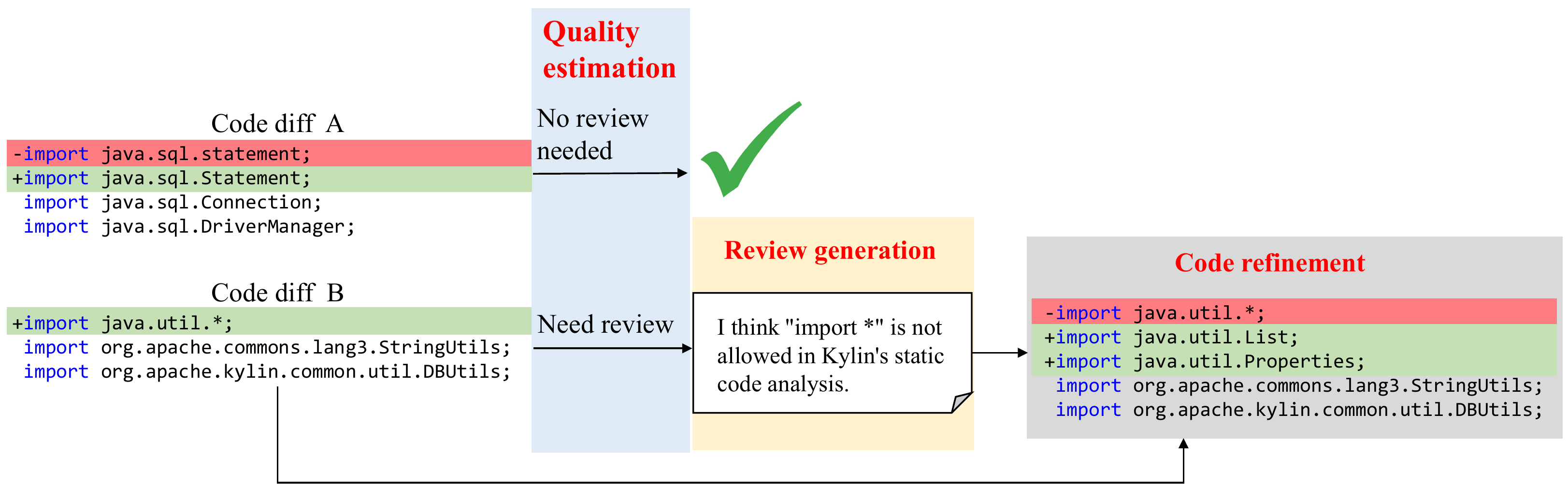}
    \caption{Overview of code review automation tasks. Three tasks are done separately.}
    \label{fig:overview_tasks}
\end{figure*}

\subsection{Code Review}
% Many contributors develop open-source projects and use collaborative software development platforms for code review. A general process of code review is shown in Figure \ref{fig:code_review}. 
In the code review process, contributors (code change author, $P_C$) update the source code to accomplish new features or fix bugs in the old version. The original code and updated code are denoted as $C_0$ and $C_1$. Once the code changes ($D: C_0 \rightarrow C_1$) are ready for review, the author of these changes creates a pull request to start the code review process.
Other peer contributors (reviewers, $P_R$) 
% \footnote{Reviewers are usually also contributors. For convenience, we use the word ``contributor'' to indicate the code change authors in most cases.}
will review the code changes and provide comments or suggestions ($R_{nl}$) on them if necessary. 
% Some review comments indicate that there exists bugs/issues in the code change. 
Based on the comments, the author makes revisions and provides a newer version of code $C_2$. We call the activities up to now as a review round. Note that the review process is not finished yet. The reviewers can further give suggestions on the revisions. And the author may also revise the code again. After a few review rounds, a pull request will finally be approved (changes are merged into the main branch) or discarded.

There are usually multiple commits in a pull request, and each commit may be related to multiple code changes across different files. Thus, to model the whole review process is difficult and challenging \cite{hellendoorn2021towards}.
% As an early step, we focus on assisting the contributors on the work in a review round.
As an early step, we focus on automating the review process of a single commit.
In the review round, two distinct roles are involved -- contributor $P_C$ who commits a code change ($D: C_0 \rightarrow C_1$), and reviewer $P_R$ who gives advice and comments ($R_{nl}$).
The goal of code review automation is to help reduce their workload. To that end, we focus on three specific code review automation tasks in both scenarios of different roles. An overview of all three tasks is shown in figure \ref{fig:overview_tasks}.

\begin{table}[t]
% \scriptsize
    % \renewcommand\arraystretch{0.8}
    \centering
    \caption{Summary of notations and symbols in this paper.}
    \label{tab:notation}
    \begin{tabular}{ll}
    \toprule
Notation & Definition \\
    \midrule
    $P_C, P_R$ & Author and reviewer of the code change \\
    $C_0, C_1, C_2$ & Different versions of source code \\
    $D$ & Code change between $C_0$ and $C_1$ \\
    $R_{nl}$ & Review comment written in natural language \\
    $\mathcal{X}, \mathcal{Y}$ & Model input and output space \\
    $c_1,c_2,\cdots,c_n$ & Source code tokens \\
    $w_1, w_2, \cdots, w_m$ & Review comment tokens \\
    $\mathcal{L}$ & Training loss function \\
% $\mathcal{D}^t,\mathcal{D}^v,\mathcal{D}^e$ & Train set, valid set \& eval set. \\
% $\mathcal{X},\mathcal{Y}$ & Source code space \& annotation space. \\
% $x=(t_1,\cdots,t_l)$ & Token sequence of the source code. \\
% $(s_1,\cdots,s_{l'})$ & Subtoken sequence of the source code. \\
% $C,\Theta_C$ & DL model \& its trainable parameters. \\
% $y, \tilde y$ & Ground truth \& mode prediction. \\
% $\mathcal{L}(\tilde y, y)$ & Loss function. \\
% $Q,K,V$ & Query, key \& value in attention. \\
% $h=(h_0,\cdots,h_{l'})$ & Context-aware hidden states. \\
% $\alpha=(\alpha_1,\cdots,\alpha_{n_h})$ & Multi-head attention with $n_h$ heads. \\
    \bottomrule
    \end{tabular}
\end{table}

\subsection{Code Change Quality Estimation}
Code change quality estimation task is to predict whether a code change is high-quality and ready to be accepted in the review process. This is a binary classification task, i.e., $\mathcal{Y} = \{0, 1\}$. The input is a code change, i.e., $\mathcal{X} = \{D(C_0, C_1)\}$.
As stated before, there are usually lots of changes across different source code files in a pull request.
It takes reviewers a large amount of time to review all the changes. 
But it always turns out that most of the changes are minor and don't need a comment or suggestion.
To improve the efficiency of code review, we define and advance to automating the code change quality estimation task.
Given the estimates, reviewers can give questionable code changes a higher priority to review and pay more attention to them, saving time and effort. Contributors can also leverage the estimates to improve low-quality code changes before submitting them to reviewers.

\subsection{Code Review Generation}
Code review generation is a sequence generation task. The output is a predicted review comment, i.e., $\mathcal{Y} = \{w_1, \cdots, w_n\}$ where $w_i$ is a natural language word and $n \in \mathcal{N}$ is the length of review comment. The input is still a code change, i.e., $\mathcal{X} = \{D(C_0, C_1)\}$, with its context.
In some previous works \cite{tufano2021towards,tufano2022using,gupta2018intelligent}, researchers use the changed code as input but not the code diff, without taking into account that review comments have to focus on the changed part. It's not recommended for reviewers to give suggestions to the code context which has not been revised.
Considering the naturalness of software\cite{hindle2012naturalness}, language models can capture general statistical properties of programs due to programs tend to be repetitive and predictable.
In the code review process,
% reviewers provide suggestions or requirements to the author in review comments. 
there are also common issues in some code changes. For example, reviewers often write ``This method should be private'' to suggest the contributor add a private decorator to a method. This gives us a chance to learn the general code review patterns and generate comments automatically to lighten the burden of reviewers.
In the real-world scenario, the model generates few review comment candidates and the reviewers may choose an ideal one from them, free from writing comments manually.

% But writing review comment is much more flexible than some sequence to sequence tasks such as neural machine translation. Because reviewers must understand the code changes and the code specifications in the project.

\subsection{Code Refinement}
In the code refinement task, the model takes as input both the original code $C_1$ written by the contributor and the review comment $R_{nl}$ from the reviewer, and targets to generate the revised version code $C_2$ implementing the requirements mentioned in $R_{nl}$.
Different from the code refinement task in other software development scenarios where only source code is taken as input \cite{tufano2019empirical}, we design specifically for the code review process to utilize the review comment as guidance for refining.
% the code refinement task is targeted to support contributors. 
Code refinement is a task designed for assisting contributors. When they open pull requests and get review comments as feedback, the model helps to revise the submitted code automatically based on the comments.

\begin{figure*}[t]
    \centering
    \includegraphics[width=0.95\textwidth]{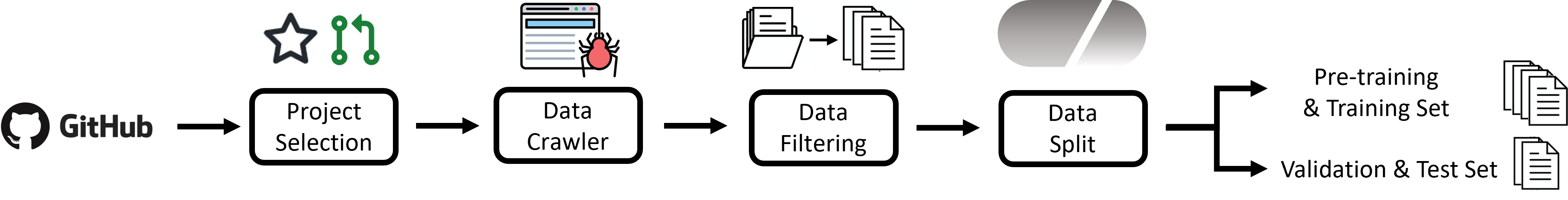}
    \caption{The process of building our dataset.}
    \label{fig:dataset}
\end{figure*}

\subsection{Data Format}
\label{sec:task:format}
All three tasks are related to source code or code changes. In the code review process, there are usually multiple related files and functions \cite{hellendoorn2021towards}. A source code file or a function/method is too large to process, because there may be multiple review comments and code revisions spreading across a file or function/method. Instead, we define the inputs of the three tasks at diff hunk level.

In pull requests, the original version and revised version of code are shown in a special format: diff (Figure \ref{fig:overview_tasks}). The diff file is generated by comparing two files before and after the change. The diff tool finds sequences of lines common to both files, interspersed with groups of differing lines called hunks. Specifically, a diff hunk is a sequence of source code lines surrounded by a few unchanged lines. Diff hunk includes the lines deleted from the original file and added to the new file. The diff format is frequently used by programmers to show the changes in the code clearly. Moreover, the diff format is an efficient representation for code changes because the unchanged lines occur only once, and the changes are aligned with ``-'' and ``+'' tags at the begging of each line. 
% The adjacent changed lines are allocated to the same hunk while distant lines are divided into different hunks. Therefore, code changes in different hunks (with long distance) are less likely to be related.
In the first two tasks, we formulate the input as a diff hunk representing the code change. In the code refinement task, we extract the input lines ($C_1$) and output lines ($C_2$) from the revision diff hunk.

\section{Code Review Dataset}

Nowadays, many developers use collaborative software development platforms like GitHub\footnote{https://github.com/} and Gerrit\footnote{https://www.gerritcodereview.com/} not only to share their code but also to perform code review activities. 
GitHub platform allows contributors to make all these review details publicly available, including the exact content of code changes, review comments and their authors, and happening time in each pull request, which makes it possible to collect and analyze code review data from open-source projects.
Hence, we build our CodeReview dataset based on pull requests data collected from open-source projects, covering nine programming languages, in GitHub.

\subsection{Project Selection}
To ensure the quality of our dataset, we collect pull request data from publicly available high-quality open-source repositories. We first sort the projects by ``popularity'' as indicated by the number of stars.
To improve the generalizability of the models trained on our dataset, we collect projects in nine of the most popular programming languages in GitHub, including C, C++, C\#, Go, Java, JavaScript, PHP, Python, and Ruby.
Then, we keep the top 10,000 projects for each of the nine programming languages and remove those projects that do not explicitly permit the re-distribution of their data. To be specific, all repositories with popular licenses such as Apache-2.0 license and MIT license \footnote{Apache-2.0, GPL-3.0, MIT, BSD-2.0, BSD-3.0, BSL-1.0, GPL-2.0 license, etc.} are kept. If the contributors write in the license file that they allow for re-distribution of their data, their projects are also collected. In order to ensure the quality of the project and acquire code review data as much as possible, we further sort the projects by the number of pull requests and filter out projects with less than 1,500 pull requests. This process keeps only active projects with many contributors and removes repositories that are forked from other projects as the pull request number is not inherited.
Then we start crawling pull request information of the projects.

\subsection{Review Data Collection}
We use GitHub REST API to collect pull requests data from the projects. GitHub API offers an HTTP request-based API for accessing repository information. By sending HTTP requests to GitHub, we can access the branches, commits, pull requests, code diff, review comments, etc in Json format conveniently. Many researchers have used GitHub API to collect and analyze pull request data in their work \cite{heum2021etcr, siow2020core}.
To advance the research of code review, \citet{heum2021etcr} develop the ETCR infrastructure for mining code review datasets from any GitHub project which practices pull-request-based development. ETCR tool can be set up requiring only a GitHub API key and repository name. The ETCR uses a Kotlin-based web crawler and works in four stages to collect data of pull requests, commits, comments and source files via the GitHub API. All the data is stored in a relational database and thus can be accessed conveniently.
We first use the ETCR tool \cite{heum2021etcr} to collect meta-information of pull requests and review comments, including the git commit hash and changed file name related to the comments. Using the meta-information,
we can further query GitHub API to get code changes (including the original file, new file, and the code diff) corresponding to the review comments. These code changes and comments make up our CodeReview data. By now, we have collected all required data for building datasets for the three downstream tasks.

\subsection{Dataset Construction}
\label{sec:dataset:constr}
The pre-training dataset is a set of code changes with or without review comments. However, it requires further data processing to build datasets for the three downstream tasks: 
\ding{182} Code change quality estimation: All commented code changes are regarded as suspicious code that introduces bugs or conflicts with code specifications. Other code changes without comments are labeled as correct. As the number of code changes without comments is about 2-3 times the number of commented code changes, we perform random down-sampling on the changes without review comments to build a balanced dataset. \ding{183} Review comment generation: Code changes with review comments will be used for this task. We filter out those comments written by the code changes author. When there is more than 1 comment related to a diff hunk, we only keep the earliest one.
\ding{184} Code refinement: We traverse each pull request in the projects. For each commented code change, we check all the commits in this pull request to find whether there is a later commit that updates this part of code again.
To be specific, when the reviewer writes a comment on the code diff $D: C_0 \rightarrow C_1$ and the revised source code lines in $C_1$ are further modified to a newer version $C_2$ in a later commit, we assume that the comments on the early change helped the contributor to make the later updates. Then the triplets $(C_1, R_{nl}, C_2)$ are collected to build the code refinement dataset. 
In some cases, a comment is related to multiple future code revisions or multiple comments contribute to a revision together. All these samples are removed from the dataset because modeling the interactions between multiple comments and multiple changes is beyond the discussion of this paper.
% Note that this not guaranteed. So we perform a human evaluation on samples from the refinement dataset. It shows that xx\% of the data are correctly aligned.
Even though we collect data from projects with high star numbers and PR numbers, the dataset quality still varies, especially when it comes to review comments. To filter out low-quality comments, we clean the pre-training and downstream datasets carefully following the steps described in Appendix (attached as supplementary material).

\subsection{Data Split}
We use the CodeReview data to build the pre-training dataset and the benchmark dataset for the three downstream tasks. To prevent the dataset from information leakage, we split the data in project level. The code repositories with more than 2,500 pull requests are used to build the pre-training dataset and training set of the benchmark dataset. Other code repositories with $[1,500, 2,500)$ pull requests are used to build the validation and test dataset.

\begin{figure*}[t]
    \centering
    \includegraphics[width=0.95\textwidth]{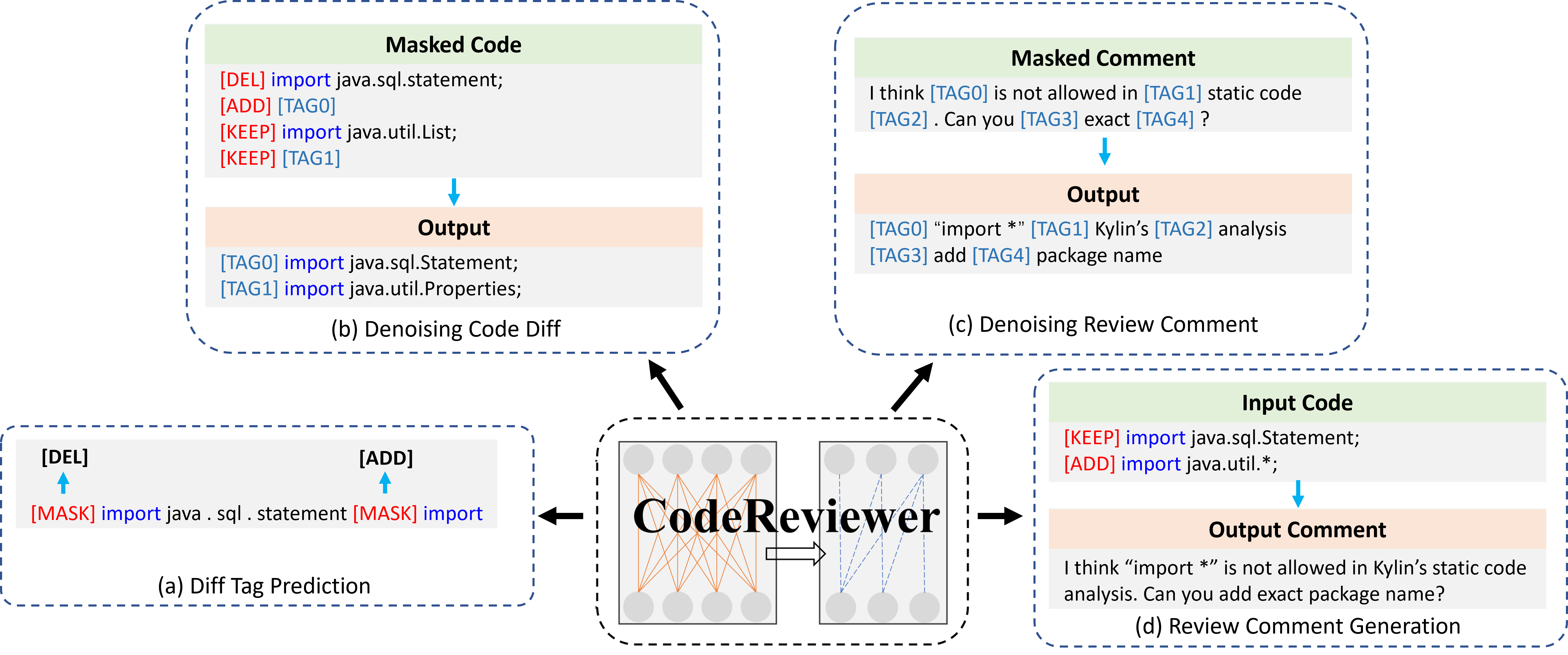}
    \caption{Pre-training tasks of CodeReviewer.}
    \label{fig:pretrain}
\end{figure*}

\section{CodeReviewer}
In this section, we describe our CodeReviewer model in detail, including the model architecture, the input-output representations of the model, and the pre-training tasks designed for code review.

We develop CodeReviewer based on Transformer \cite{vaswani2017transformer} 
% We build a Transformer \cite{vaswani2017transformer} based encoder-decoder model 
and design four pre-training tasks related to the code review process to improve the model's capacity for automating code review activities.

\subsection{Model Architecture}
% CodeReviewer is a Transformer \cite{vaswani2017transformer} based encoder-decoder model. 
The CodeReviewer is an encoder-decoder model based on Transformer \cite{vaswani2017transformer}.
% We follow the work of T5 \cite{raffel2019exploring} and CodeT5 \cite{wang2021codet5}, and use the exact same architecture as T5-base and CodeT5-base. 
We adopt the same architecture as Text-To-Text-Transfer Transformer (T5) model \cite{raffel2019exploring}.
The CodeReviewer model consists of 12 Transformer encoder layers and 12 decoder layers. There are 12 attention heads in each layer and the hidden size is 768. The total parameter size of the model is 223M.

We initialize CodeReview with the parameters of CodeT5 \cite{wang2021codet5}. We further pre-train the model with our four pre-training tasks. Once pre-trained, CodeReviewer is finetuned and evaluated on the downstream tasks respectively.

\subsection{Input-Output Representations}
CodeReviewer takes different inputs and outputs for different tasks.
For code understanding tasks or comment generation tasks, the model takes code diff as input. For the code refinement task, the model takes as input both the original source code and the review comment.

Following the standard way of input processing in Transformer, the input is treated as a token sequence. We use the same RoBERTa \cite{liu2019roberta} tokenizer as CodeT5 to split the source code and review comment to the tokens. A special token $[CLS]$ is prepended to the sequence, forming the input as $\{[CLS], c_1, c_2, \cdots, c_n\}$ where $c_i$ is source code token and $n$ is the sequence length. To help our model understand the diff format better, the special line tags ``-'' and ``+'' in diff file indicating line deletion and line insertion are replaced with special tokens $[DEL]$ and $[ADD]$. We also insert a $[KEEP]$ before each unchanged line. When there are both source code tokens and review comment tokens in the input, a $[MSG]$ token is inserted to separate the two sequences. Thus the input is $\{[CLS], c_1, c_2, \cdots, c_n, [MSG], w_1, w_2, \cdots, w_m\}$.

The model outputs consist of two components: token representations from the encoder and generated token sequence from the decoder. For classification tasks, the representation of $[CLS]$ token is used to generate the predictions. For sequence generation tasks, the decoder outputs the predicted token sequence. 
% Similar to the input processing, the target outputs are also split into a token sequence by RoBERTa tokenizer.

\subsection{Pre-training Tasks}
\label{sec:pret}
The key challenge of automating code review activities is to understand code changes and capture the relationship between code changes and corresponding review comments. Therefore, we design four pre-training tasks to improve the abilities of CodeReviewer.
% There are three main challenges in the automation of code review: \ding{182} Understanding the code changes. This is essential for all three tasks. \ding{183} Learning the distribution of review comments, which requires both understanding and generating review comments for code refinement and comment generation respectively. \ding{184} Modeling the relationship between code changes and corresponding review comments. This is essential for our code review model to make comments given a code change and to update code based on review comments.

\subsubsection{Diff Tag Prediction}

%  prevent from duplicates in old and new version
As mentioned in Section \ref{sec:task:format}, diff is a special format containing rich information in code changes.
% From a diff string we can directly figure out the deleted, inserted, and unchanged lines. 
Compared with encoding both the original and new version code, the diff format prevents duplicates of unchanged lines. Thus, learning code diff as a special format of source code is critical for the model in the code review scenario.

We use Diff Tag Prediction (DTP) task to equip the model with the ability to understand the special line tags in code diff. Note that we replace ``+'' and ``-'' tags at the beginning of diff lines by special tags $[ADD]$, $[DEL]$ and involve $[KEEP]$ in the model input. In DTP training, the three kinds of special tags are replaced by $[MASK]$ tokens in the model input. We train our model to predict special tokens $[ADD]$, $[DEL]$ or $[KEEP]$ at corresponding positions. By training on DTP, the model learns to distinguish whether a line is unchanged or updated. This helps CodeReviewer to understand the code diff format. Concretely, model predicts a probability distribution $\mathbf{p}^{(i)}=(p_0^{(i)}, p_1^{(i)}, p_2^{(i)})$ for the $i$-th masked tag position and a standard cross-entropy loss is used to train our model:

\begin{equation}
    \mathcal{L}_{DTP} = -\sum_{i} \left( y^{(i)}_0\log p^{(i)}_0 +  y^{(i)}_1\log p^{(i)}_1 +  y^{(i)}_2\log p^{(i)}_2  \right)
    \label{eq:tag}
\end{equation}

\subsubsection{Denoising Objective}

Denoising objective is an unsupervised pre-training objective introduced by \citet{lewis2019bart} to acquire a general understanding of a corpus. 
% In the original denoising objective, model is fed with a sentence with 15\% of original tokens masked randomly and is trained to predict them. 
The original denoising objective randomly masks spans in the input sentence with a corrupted rate of 15\% and predicts these spans.
By learning to predict masked spans, the model gains general knowledge about the corpus distribution. 
% Model pre-trained with denoising objective can be used for both classification and generation tasks related to the corpus.

To better understand code diff and review comments, we design two denoising objectives for CodeReviewer: denoising code diff (DCD) and denoising review comment (DRC). In the DCD task, we randomly select 15\% code lines and mask them. We corrupt inputs on line-level but not span-level in order to keep the integrity format of code diff. It's worth noting that to reduce the difficulty of this task, the line tags ($[ADD]$, $[DEL]$ and $[KEEP]$) are preserved to inform the model whether a line is updated or removed. The DCD task aims to help the model learn the distribution of code changes, augmenting both the encoder and decoder. Encoder producing a better representation of code changes benefits both understanding and generation tasks. Pre-training the decoder on DCD helps the model generate better source code in the code refinement task.
In DRC task, the model is given a corrupted review comment as input and is required to recover the masked spans. Specifically, we randomly mask spans with 20\% corrupted rate and train the decoder to generate them. Training with the DRC task is expected to benefit comment-related tasks. We describe the loss of DCD and DRC tasks as:
\begin{equation}
    \label{eq:diff}
    \mathcal{L}_{DCD} = \sum_{t=1}^k -\log P_\theta(c_t
    | \mathbf{c}^{\text{mask}},\mathbf{c}_{<t})
\end{equation}
\begin{equation}
    \label{eq:msg}
    \mathcal{L}_{DRC} = \sum_{t=1}^k -\log P_\theta(w_t
    | \mathbf{w}^{\text{mask}},\mathbf{w}_{<t})
\end{equation}
where $\mathbf{c}^{\text{mask}}, \mathbf{w}^{\text{mask}}$ are masked code diff and masked review comment, and $\mathbf{c}_{<t}, \mathbf{w}_{<t}$ are the span sequence of code diff and review comment generated so far.

\subsubsection{Review Comment Generation}
In each pre-training task mentioned above, only one modal is involved, which means that the model only learns to understand source code or review comments in one task. However, the most challenging part of code review is to capture the relationship between code changes and review comments. To equip our model with this capability, we utilize bi-modal data (code changes in programming languages and related review comments in natural language) in the pre-training task. For generalizability, we use a simple conditional generation task: review comment generation (RCG). In RCG, the model is given a code change as input and asked to generate the review comment written by the human reviewer. We use the negative log-likelihood loss:
\begin{equation}
    \mathcal{L}_{RCG} = \sum_{t=1}^k -\log P(w_t|\mathbf{c},\mathbf{w}_{<t})
\end{equation}
where $\mathbf{c}$ is code change and $\mathbf{w}_{<t}$ is the comment generated so far.

\subsection{Fine-Tuning}
We group all downstream tasks into classification tasks and generation tasks. For classification tasks such as code change quality estimation, we only use the pre-trained encoder. The representation in the last layer of the special token $[CLS]$ at the beginning $h_0$ is fed into a linear classifier to produce the prediction.
For generation tasks such as code refinement, the entire pre-trained encoder-decoder model is used. A standard token negative log-likelihood loss is leveraged to optimize the probability of the target sequence.

\section{Study Design}
To investigate the performance of our CodeReviewer model on the code review tasks, we perform a large-scale study to answer the following research questions (RQ):
\paragraph{RQ1: How does CodeReviewer perform on the code change quality estimation task?} We provide as input to CodeReviewer a code change $D_1: C_0 \rightarrow C_1$
% represented in diff format
, and ask the model to give a binary prediction whether the code change needs a review.

% \paragraph{RQ2: To what extent is CodeReviewer able to generate review comments automatically as reviewers would do?} 
\paragraph{RQ3: How does CodeReviewer perform on the review generation task?}
In this RQ, CodeReviewer is also provided a code change $D_1: C_0 \rightarrow C_1$, but we assess the ability of CodeReviewer to generate a natural language comment $R_{nl}$ as the reviewers would do.

\paragraph{RQ3: How does CodeReviewer perform on the code refinement task?} Given a few lines of source code submitted for code review and the feedback from reviewers written in natural language, the model is required to refine the submitted code to implement the requirements of the review comments.

In RQ1-RQ3, we evaluate CodeReviewer on different code review tasks. To assess the contribution of each pre-training task and our multilingual dataset, we further investigate RQ4 and RQ5.

\paragraph{RQ4: What role does each pre-training task play in CodeReviewer?} In RQ1-RQ3, the evaluated CodeReviewer model is pre-trained on all four pre-training tasks before being applied to the downstream tasks. We remove the pre-training tasks one by one and evaluate the resulting models again to expose the influence of each pre-training task on different downstream tasks.

\paragraph{RQ5: Can multilingual dataset benefit model performance on understanding single programming language?} Existing research show that multilingual training dataset can benefit for model performance compared with monolingual dataset on neural machine translation and code translation \cite{chiang2021multinmt, zhu2022multilingual}, especially for low-resource languages. We evaluate the performance of CodeReview pre-trained and fine-tuned on monolingual dataset and compare it with the model trained on the full dataset to show the influence of multilingual dataset.

\paragraph{Dataset}
We build our pre-training dataset and 3 downstream task datasets as described in Section \ref{sec:dataset:constr}.
Table \ref{tab:sta_pret} summarizes the statistics of the pre-training dataset. For benchmark datasets, details are shown in Table \ref{tab:sta_bench}.

\begin{table}[t] \footnotesize
    \centering
    % \fontsize{8}{11}\selectfont
    % \renewcommand\arraystretch{0.9}
	\caption{Statistics of pre-training dataset.}
	\label{tab:sta_pret}
	\begin{threeparttable}
		\begin{tabular}{lccccc}
			\toprule
			\multirow{2}{*}{Language} & \multicolumn{3}{c}{Meta Info} & \multicolumn{2}{c}{Data \#} \\			\cmidrule(r){2-4} \cmidrule(r){5-6} &
			Project & PRs\tnote{\textdagger} & Data Size & w/o comment & w/ comment \\
			\midrule
			Python & 195 & 1,451k & 72.8G & 887k & 518k \\
            Java  & 175 & 1,073k & 54.8G & 876k & 467k \\
            Go & 146 & 951k & 40.4G & 728k & 410k \\
            C++ & 133 & 999k & 82.1G &  474k & 202k \\
            JavaScript  & 194 & 1,354k & 30.6G &  425k & 293k \\
            C  & 77 & 441k  & 135.4G & 292k & 110k \\
            C\#  & 77  & 463k  & 28.2G & 324k & 199k \\
            Php & 92 & 574k & 16.0G & 215k & 157k \\
            Ruby & 72 & 626k & 3.8G & 90k  & 126k \\
			\midrule
			Total & 1,161 & 7,933k & 463.2G & 4,311k & 2,481k \\
            \bottomrule
		\end{tabular}
 		\begin{tablenotes}
            \item [\textdagger] Pull request numbers of the projects in total.
        \end{tablenotes}
	\end{threeparttable}
\end{table}

\begin{table}[t] \small
    \centering
    % \fontsize{8}{12}\selectfont
    % \renewcommand\arraystretch{0.9}
	\caption{Statistics of benchmark datasets.}
	\label{tab:sta_bench}
	\begin{threeparttable}
		\begin{tabular}{lcccc}
			\toprule
			Dataset & Train \# & Valid \# & Test \# & LOC \\
			\midrule
            % Diff Quality & \multirow{2}{*}{~266k} & \multirow{2}{*}{~31k} & \multirow{2}{*}{~31k} & \multirow{2}{*}{~11M} \\
            % Estimation \\
            % Review Comment & \multirow{2}{*}{~118k} & \multirow{2}{*}{~10k} & \multirow{2}{*}{~10k} & \multirow{2}{*}{~1.8M} \\
            % Generation \\
            Quality Estimation & $\sim$266k & $\sim$31k & $\sim$31k & $\sim$11M \\
            Review Comment Generation & $\sim$118k & $\sim$10k & $\sim$10k & $\sim$1.8M \\
            Code Refinement & $\sim$150k & $\sim$13k & $\sim$13k & $\sim$1.3M \\
            \bottomrule
		\end{tabular}
	\end{threeparttable}
\end{table}

\subsection{Baseline Models}
To demonstrate the superiority of our multilingual code review related pre-training dataset and carefully designed pre-training tasks, we compare our CodeReviewer model with three baselines, including a state-of-the-art (SOTA) model architecture Transformer \cite{vaswani2017transformer} trained from scratch and two pre-trained models: T5 for code review \cite{tufano2022using} and CodeT5 \cite{wang2021codet5}.

\paragraph{Transformer.} Transformer \cite{vaswani2017transformer} is a SOTA model architecture for many classification and generation tasks. The key component of Transformer is the multi-head attention module and the parameterized linear transformation layers. We use the same model setting as CodeT5-base, with a 12-layer encoder and a 12-layer decoder.

\paragraph{T5 for code review.} \citet{tufano2022using} attempt to use pre-trained model for code review automation. 
% They collect a pre-training dataset including both source code and corresponding English description from two sources: Stack Overflow dump and CodeSearchNet \cite{husain2019codesearchnet}. 
They pre-train a small version of T5 model with the denoising objective proposed by \citet{raffel2019exploring} on their own dataset including Stack Overflow dumps and CodeSearchNet \cite{husain2019codesearchnet}.
% and use it for downstream code review tasks.
The model consists of 61M parameters, with 6 layers for both encoder and decoder. We refer to this model with T5 for convenience later.

\paragraph{CodeT5-base.} CodeT5 is a SOTA unified pre-trained encoder-decoder model for code understanding and generation tasks proposed by \citet{wang2021codet5}.
They propose to pre-train CodeT5 with identifier-aware denoising tasks and bimodal dual generation objective, which makes CodeT5 the SOTA model for multiple code-related downstream tasks including code summarization, code generation, code translation and code refinement.
The base version of CodeT5 consists of 12 encoder layers and decoder layers with the parameter size 220M. We use CodeT5 to refer to this model.

\subsection{Evaluation Metrics}
\label{sec:metric}
We provide a brief description of the evaluation metrics used for the three downstream tasks in this section.

\paragraph{Code change quality estimation.} It is a binary classification task. We use accuracy, precision, recall, and F1 to evaluate the model predictions. Note that when computing the later 3 metrics, the code changes with issues (requires for comments and updates) are treated as the positive class.

\paragraph{Review comment generation.} We compute the BLEU (Bilingual Evaluation Understudy) \cite{papineni2002bleu} score of the predictions. BLEU is widely used to assess the quality of automatically generated text in generation tasks such as machine translation and code generation. We use the BLEU-4 variant, which computes the overlap of $n$-grams between the generated text and the reference text with $n$ from 1 up to 4. As review comments are diverse and non-unique, we further apply human evaluation on generated comments from two perspectives: information and relevance.
% how much information the generated review comment conveys and to what extent the comment is related to the code change. 
% The evaluation standard is described in Section \ref{exp:rq2} in detail.
\ding{182} Information: in this metric, we evaluate how informative the comment is for the contributor to revise the source code. Comments like ``declare this variable as private'' are more informative than those like ``why do we need this?''. \ding{183} Relevance: in this metric, we evaluate to what extent the review comment is related to the corresponding code change. The comments point out the issues in the code changes will get high scores, while those not related or disobey the logic in the code changes should get low scores. The comments are labeled with a 1-5 score for each metric. We describe in detail of the two metrics in the Appendix.

\paragraph{Code refinement.} For this task, we compute the BLEU score between generated code and target code and the exact match (EM) rate. BLEU only evaluates the similarity between the prediction and the target, while a small change in source code can result in compile error and execution error. So exact match is a more important metric in this task. Only if a prediction is exactly the same as the target, will it be considered as correct.

\subsection{Implementation Details}
We implement our model with the popular deep learning development framework PyTorch \footnote{https://pytorch.org/} and the python package transformers developed by HuggingFace \footnote{https://huggingface.co/}.
We pre-train our CodeReviewer model with 2 DGX-2 servers with 16 NVIDIA V100-32G GPUs on each server. The learning rate and batch size in the pre-training stage are set to 0.0002 and 768. We use AdamW optimizer with linear warmup to optimize the model for 250k steps. The warmup step number is set to 2500. When fine-tuning our CodeReivewer and baseline models on the three downstream tasks, we use batch size 72 and learning rate 0.0003. Each experiment on downstream tasks is performed on a server with 4 V100 GPUs. When evaluating the review comment generation task and the code refinement task, we use beam search with size 10.

\section{Results Analysis}
In this section, we answer each research question based on our experiment results. We first focus on three research questions concerning our model's performance in each task and how it compares to other state-of-the-art baseline models. We further demonstrate the effects of our pre-training tasks and multilingual dataset.

\subsection{RQ1: Performance on Code Change Quality Estimation}

Table \ref{tab:res_cls} shows the results on the code change quality estimation task. From the table, we can see that whether the baselines are trained from scratch or pre-trained models, CodeReviewer outperforms them significantly on all four metrics. Specifically, our CodeReviewer model improves F1 and accuracy by 8.24\% and 7.07\% compared with T5. The improvement over CodeT5 is also over/about 7\%, which demonstrates that our pre-training tasks help CodeReviewer understand code changes better. Besides, the performance of Transformer trained from scratch is inferior to the other three models, indicating the importance of pre-training.

\begin{table}[t] % \scriptsize
    \centering
    % \fontsize{8}{12}\selectfont
    % \renewcommand\arraystretch{0.9}
	\caption{Results on code change quality estimation.}
	\label{tab:res_cls}
	\begin{threeparttable}
		\begin{tabular}{lcccc}
			\toprule
			 Model (Layers \#) & Precision & Recall & F1 & Accuracy \\
			\midrule
			Transformer (12) & 74.50 & 46.07 & 56.93 & 65.16 \\
			T5 (6) & 70.82 & 57.20 & 63.29 & 66.82 \\
			CodeT5 (12) & 70.36 & 58.96 & 64.16 & 67.07 \\
			\midrule
			CodeReviewer (12) & \textbf{78.60} & \textbf{65.63} & \textbf{71.53} & \textbf{73.89} \\
            \bottomrule
		\end{tabular}
	\end{threeparttable}
\end{table}

\subsection{RQ2: Performance on Review Generation}
\begin{figure*}[t]
    \centering
        \includegraphics[width=0.95\textwidth]{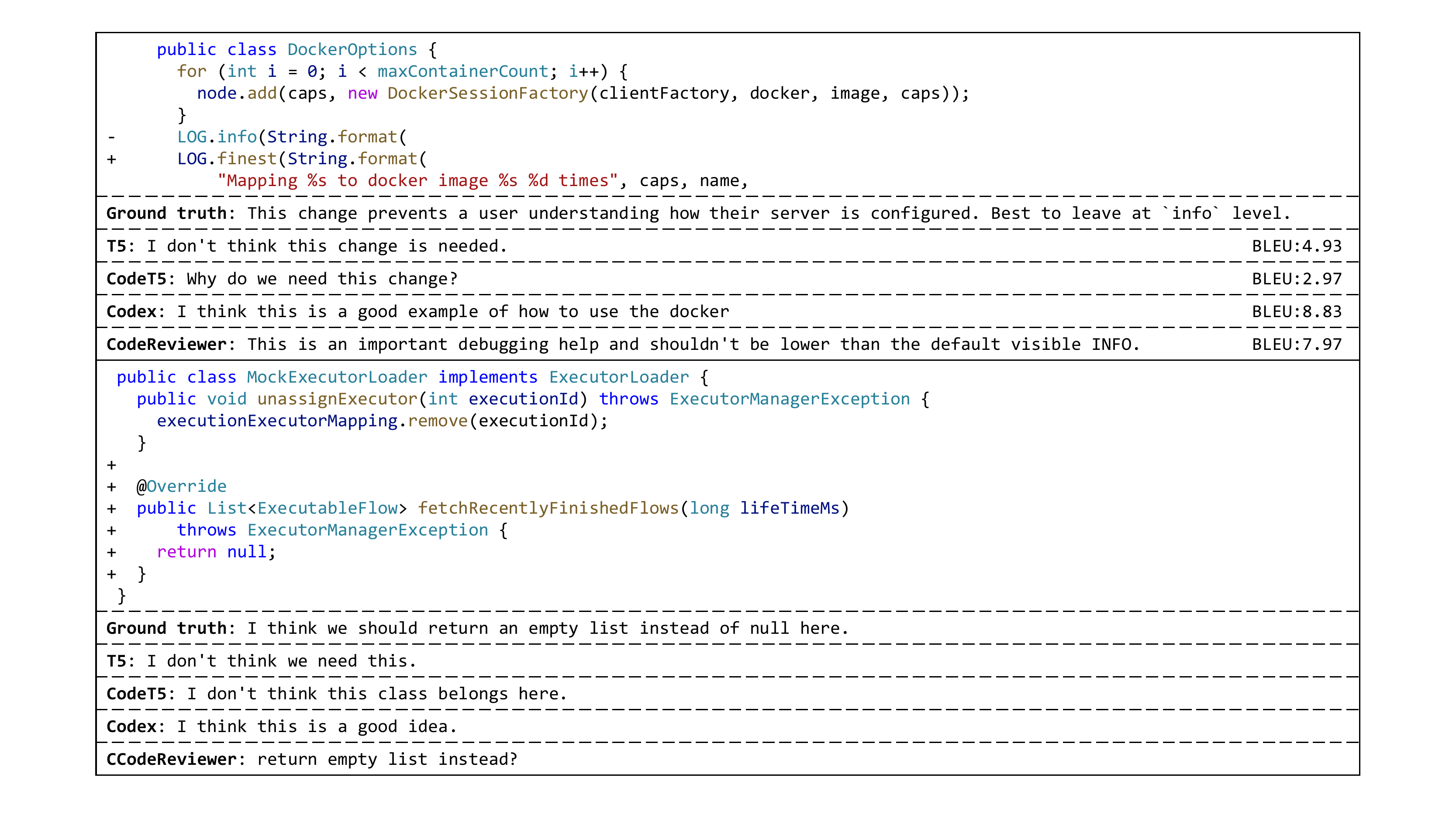}
        \caption{Two examples of the review generation task. The Codex output is obtained by Copilot.}
        \label{fig:ex_cmt}
\end{figure*}

Table \ref{tab:res_gen} shows the results on the review generation task. CodeReviewer produces a higher BLEU score than the baseline models. However, the BLEU score of our model is still lower than 10, indicating it is a hard task. As mentioned before, review comments are diverse and non-unique. Referring to the example illustrated in Figure \ref{fig:ex_cmt}, our model prediction conveys similar intent as the ground truth. But their words differ greatly so that the BLEU score is lower than Codex. To investigate the models' performance better, we perform a human evaluation on their predictions.

We firstly choose 300 random samples in Java and Python from the test set and manually select 100 of them with high-quality code changes and review comments. For the selected samples, we invite 6 professional sophisticated programmers to score the reference comment and generated comments of each model as described in Section \ref{sec:metric}. Results are listed in Table \ref{tab:res_gen}. Our model improves the information score and relevance score of the generated comments by about 20\% relatively compared with the baseline models, which means comments generated by CodeReviewer are more useful and adequate. The performance of T5 is inferior to the other models, including the Transformer baseline. Because the parameter size of T5 (61M) is about 1/4 of the other models (223M). We conjecture that it is too challenging for a small model to generate the review comments.

In this task, we also conduct a qualitative analysis of the world's largest language model on source code, Codex \cite{chen2021evaluating}, which has been proven to be the most powerful code generation model and capable of zero-shot task transfer. Since we don't get access to the model and can't fine-tune it, we alternately use GitHub Copilot \cite{copilot} which is powered by Codex. By providing several code diff and review comment pairs as prompt, we find that Codex can't generate any meaningful reviews but copying comments in the given examples. Figure \ref{fig:ex_cmt} shows two examples of outputs generated by different models including Codex.

\begin{table}[t] % \scriptsize
    \centering
    % \fontsize{8}{12}\selectfont
    % \renewcommand\arraystretch{0.9}
	\caption{Results on review comment generation. The perfect scores of two metrics in human evaluation set are both 5.}
	\label{tab:res_gen}
	\begin{threeparttable}
		\begin{tabular}{lccc}
			\toprule
			\multirow{2}{*}{Model  (Layers \#)} & Test set & \multicolumn{2}{c}{Human evaluation set} \\
			\cmidrule(r){2-2} \cmidrule(r){3-4}
			& BLEU & Information & Relevance \\
			\midrule
			Transformer (12) & 4.76 & 3.08 & 2.50 \\
			T5 (6) & 4.39 & 2.54 & 1.62 \\
			CodeT5 (12) & 4.83 & 3.03 & 2.40 \\
			\midrule
			CodeReviewer (12) & \textbf{5.32} & \textbf{3.60} & \textbf{3.20} \\
            \bottomrule
		\end{tabular}
	\end{threeparttable}
\end{table}

% [3.08       3.02666667 2.54       3.60333333] [2.50333333 2.40333333 1.62       3.20333333]

% \begin{figure*}[t]
%     \centering
%     \begin{subfigure}{\textwidth}
%         \includegraphics[width=0.95\textwidth]{figs/task2_ex.pdf}
%         \caption{An example of the review generation task. The Codex output is obtained by Copilot.}
%         \label{fig:ex_cmt}
%     \end{subfigure}
%     \begin{subfigure}{\textwidth}
%         \includegraphics[width=0.95\textwidth]{figs/task3_ex.pdf}
%         \caption{An example of the code refinement task. To make clear, we show the code diff between the inputs and model outputs instead of their original form.}
%         \label{fig:ex_refine}
%     \end{subfigure}
%     \caption{Examples of the review generation task and the code refinement task.}
%     \label{fig:examples}
% \end{figure*}

\begin{figure*}[h]
        \includegraphics[width=0.85\textwidth]{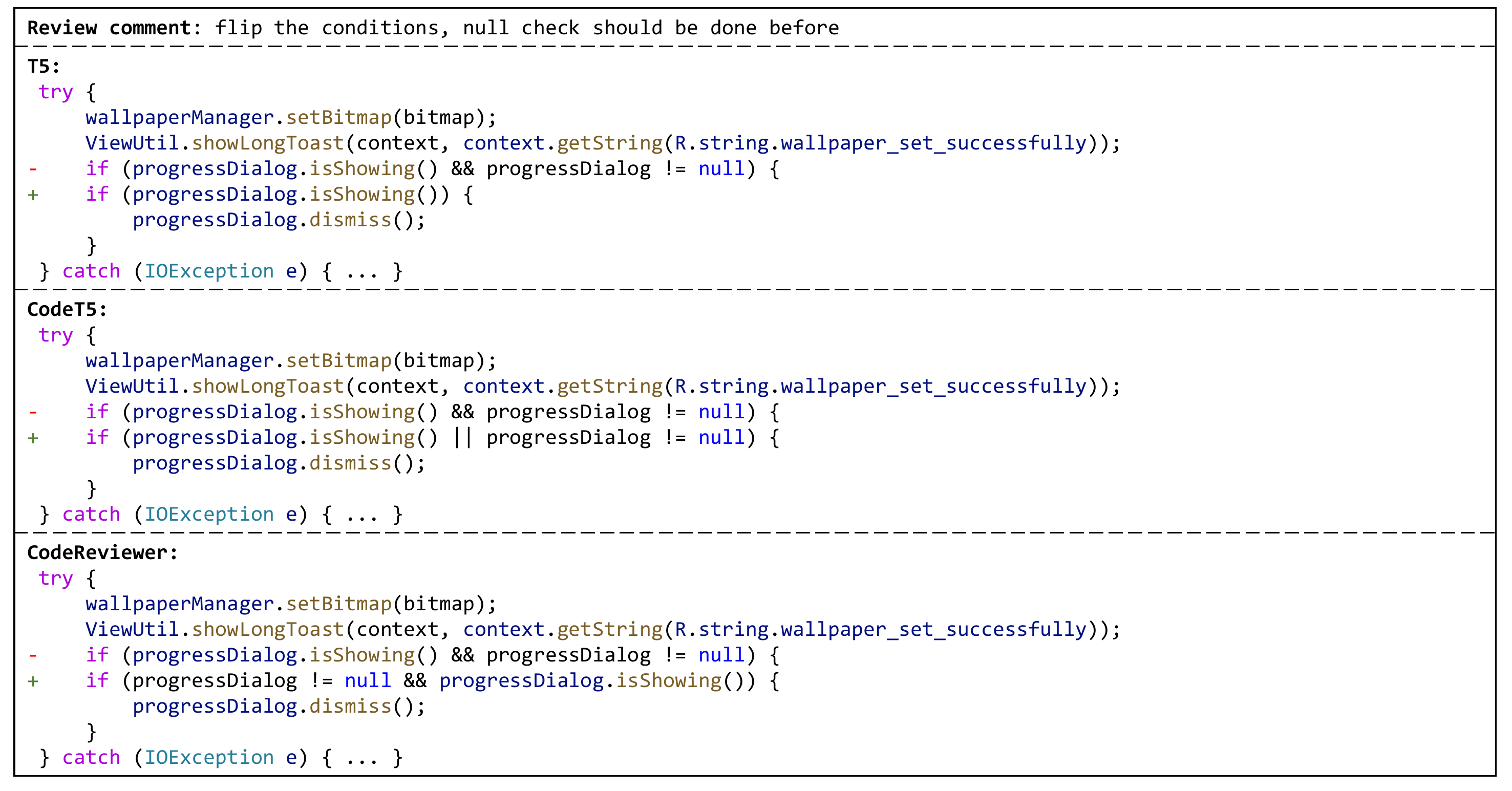}
        \caption{An example of the code refinement task. To make clear, we show the code diff between the inputs and model outputs instead of their original form.}
        \label{fig:ex_refine}
\end{figure*}

\subsection{RQ3: Performance on Code Refinement}
Table \ref{tab:res_ref} shows the results on the code refinement task. The Na\"iveCopy method directly copies the input code as the refinement result. It produces a decent BLEU score but 0.00 Exact Match (EM). 
% So BLEU score is not an adequate metric for this task and we focus on the EM score.
CodeReviewer successfully generates the repaired code exactly the same as ground truth for more than 30\% cases, which is two times as the result of T5 and 25\% more than CodeT5 relatively, demonstrating the superior ability of CodeReviewer to understand the review comments and refine code based on them. The BLEU score of CodeReviewer is also higher than T5 and CodeT5. Figure \ref{fig:ex_refine} shows an example of a perfect prediction generated by our model.
Note that the Transformer model failed to converge in 10 epochs, producing about 0 BLEU score, so the result is not listed in the table.

\begin{table}[t] % \scriptsize
    \centering
    % \fontsize{8}{12}\selectfont
    % \renewcommand\arraystretch{0.9}
	\caption{Results on code refinement.}
	\label{tab:res_ref}
	\begin{threeparttable}
		\begin{tabular}{lccc}
			\toprule
			 Model (Layers \#) & BLEU & EM \\
			\midrule
			Na\"iveCopy & 58.75 & 0.00 \\
			T5 (6) & 77.03 & 15.08 \\
			CodeT5 (12) & 80.82 & 24.41 \\
			\midrule
			CodeReviewer (12) & \textbf{82.61} & \textbf{30.32} \\
            \bottomrule
		\end{tabular}
	\end{threeparttable}
\end{table}

\subsection{RQ4: Influence of Pre-training Tasks}
We demonstrate the superior performance of CodeReviewer on different downstream tasks in RQ1-RQ3, proving that the well-designed pre-training tasks benefit for automating code review activities. In this RQ, we dive deeper and expose the contribution of each pre-training task. We further pre-train three models, each of which is trained with some pre-training tasks removed, and evaluate their performance on the code change quality estimation task. The models CodeReviewer w/o DTP, CodeReviewer w/o DCD and CodeReviewer w/o CMT represent CodeReviewer trained without the Diff Tag Prediction task, the Denoising Code Diff task, and the other two comment related tasks (Denoising Review Comment and Review Comment Generation) respectively.

% Table \ref{tab:abl_cls} and Table \ref{tab:abl_ref} show the results of this ablation study. 
Table \ref{tab:abl_cls} shows the results. 
The drop of the model performance demonstrates the importance of the pre-training tasks.
For the code change quality estimation task, all the pre-training tasks help our model understand the code diff better. But the diff tag prediction task and the denoising code diff task are more important. The results are consistent with our motivation in Section \ref{sec:pret}. 
% For the code refinement task, the two review comment related tasks and the denoising code diff tasks make the most contributions. Because code refinement task requires the model to understand the code review in the input and generate the revised code simultaneously. Note that when the diff tag prediction task is removed, the model actually produce higher BLEU score and EM score. The reason is that the input of this task is plain source code rather than code diff, while the DTP task is designed for code diff specially.

\begin{table}[t] % \scriptsize
    \centering
    % \fontsize{8}{12}\selectfont
    % \renewcommand\arraystretch{0.9}
	\caption{Ablation study on code change quality estimation.}
	\label{tab:abl_cls}
	\begin{threeparttable}
		\begin{tabular}{lcccc}
			\toprule
			 Model & Precision & Recall & F1 & Accuracy \\
			\midrule
			CodeReviewer & 78.60 & 65.63 & 71.53 & 73.89 \\
			\ \ w/o CMT & 77.14 & 66.35 & 71.34 & 73.35 \\
			\ \ w/o DTP & 79.59 & 62.64 & 70.10 & 73.29 \\
			\ \ w/o DCD & 79.87 & 60.94 & 69.13 & 72.80 \\
            \bottomrule
		\end{tabular}
	\end{threeparttable}
\end{table}

% \begin{table}[t] % \scriptsize
%     \centering
%     % \fontsize{8}{12}\selectfont
%     % \renewcommand\arraystretch{0.9}
% 	\caption{Ablation study on code refinement task.}
% 	\label{tab:abl_ref}
% 	\begin{threeparttable}
% 		\begin{tabular}{lcc}
% 			\toprule
% 			 Model & BLEU & EM \\
% 			\midrule
% 			CodeReviewer & 82.61 & 30.32 \\
% 			\ \ -ENC & 83.08 & 31.03 \\
% 			\ \ -CMT & 82.58 & 29.95 \\
% 			\ \ -DEC & 82.69 & 29.27 \\
%             \bottomrule
% 		\end{tabular}
% 	\end{threeparttable}
% \end{table}

\subsection{RQ5: Influence of Multilingual Dataset}
In the previous experiments, CodeReviewer is trained on the datasets consisting of nine programming languages. To investigate whether multilingual datasets help our model better understand a single programming language, we build monolingual datasets for Java, C\# and Ruby language, repectively. Java represents popular languages and Ruby represents the low-resource languages as shown in Table \ref{tab:sta_pret}. For each of the three languages, we pre-train and finetune the CodeReviewer on the monolingual datasets and compare the performance on the code change quality estimation task with the original CodeReviewer pre-trained and finetuned on full multilingual datasets. The results are listed in Table \ref{tab:abl_lang}.

The multilingual CodeReviewer outperforms the three monolingual models consistently, improving the accuracy by 2.32\% and the F1 score by 1.10\% on average. We conclude that our multilingual dataset benefits the CodeReviewer for understanding specific languages significantly. This reveals the superiority of our multilingual dataset over the datasets for single programming language. It also proves the broad applicability of CodeReviewer in different programming languages.

\begin{table}[t] % \scriptsize
    \centering
    % \fontsize{8}{12}\selectfont
    % \renewcommand\arraystretch{0.9}
	\caption{Ablation study of multi-lingual dataset on the code change quality estimation task.}
	\label{tab:abl_lang}
	\begin{threeparttable}
		\begin{tabular}{lcccccc}
			\toprule
			\multirow{2}{*}{Metric} & \multicolumn{2}{c}{Java} & \multicolumn{2}{c}{C\#} & 
			\multicolumn{2}{c}{Ruby} \\
			\cmidrule(r){2-3} \cmidrule(r){4-5} \cmidrule(r){6-7}
			& Multi & Single & Multi & Single & Multi & Single \\
			\midrule
            Accuracy & \textbf{74.04} & 72.45 & \textbf{74.80} & 72.21 & \textbf{82.70} & 79.92 \\
            F1 & \textbf{70.53} & 69.34 & \textbf{76.52} & 75.53 & \textbf{89.23} & 88.10 \\
            \bottomrule
		\end{tabular}
	\end{threeparttable}
\end{table}

\section{Related Works}
\subsection{Pre-training for Code-related Tasks}
Deep Learning techniques have been widely adopted in software engineering research \cite{watson2020systematic,devanbu2020deep}. 
In recent years, motivated by the great impact of pre-training technique in natural language (NL) processing area \cite{devlin2018bert,raffel2019exploring,lewis2019bart,brown2020language}, researchers have made attempts to investigate whether pre-trained models for programming language (PL) can further facilitate software development.

Since pre-trained models can address several downstream tasks by fine-tuning, previous works target at different scopes of code-related tasks with their models.
\citet{kanade2020learning} pre-train CuBERT in a Python code corpus mostly for classification tasks like variable-misuse classification, wrong binary operator, etc. 
CodeBERT \cite{feng2020codebert} and GraphCodeBERT \cite{guo2021graphcodebert} are bi-directional transformer models pre-trained on NL-PL pairs in six programming languages, the latter introduces new pre-training tasks designed for source code data flow. 
% \textsc{SynCoBERT}
Both models have shown effectiveness on code-text crossing tasks like NL code search and code summarization, and other code understanding tasks like code clone detection and defect detection thanks to bi-directional attention mechanism. 
CodeGPT \cite{lu2021codexglue}, GPT-C \cite{svyatkovskiy2020intellicode} and Codex \cite{chen2021evaluating} focus on generative tasks like code completion and code generation because they are pre-trained with decoder-only transformer architecture.
As for encoder-decoder models like PLBART \cite{ahmad2021unified} and CodeT5 \cite{wang2021codet5} and unified models like UniXcoder \cite{guo2022unixcoder}, they can be applied for both understanding and generation tasks that take source code or text as inputs. 

All the above models pay no attention to tasks in the code review process. A distinct feature of code review tasks is that code changes should be considered as inputs. Recently, \citet{tufano2022using} pre-trained a T5 model for automating code review activities. However, their work is different from ours on two points. First, they just use the pre-training objectives of T5 \cite{raffel2019exploring} and don't take into consideration how to integrate code changes into pre-training. Second, their pre-training data is not directly related to code review, but we collect data from GitHub pull requests for pre-training.

\subsection{Automating Code Review Activities}
As indicated by previous empirical studies, code review is an important part in the software development lifecycle and involves a significant amount of effort and time of reviewers \cite{sadowski2018modern, bosu2013impact}.
Researchers are paying more attention to automating code review activities, including reviewer recommendation \cite{thong2015whoshould, zanj2015peerreview}, comment location prediction \cite{shi2019automatic, hellendoorn2021towards}, review comment recommendation \cite{gupta2018intelligent, siow2020core} and code refinement \cite{tufano2021towards}.

\citet{thong2015whoshould} reveal that 4\%-30\% of reviews have code reviewer assignment problems. Thus, they propose a file location-based tool RevFinder to recommend appropriate code reviewers. To tackle the same problem,  \citet{zanj2015peerreview} design the system cHRev which utilizes code review histories to recommend reviewers for a code change. While these researchers aim at improving code review at an early stage, others are devoted to solving the more challenging tasks during the code review process. \citet{shi2019automatic} propose the DACE framework based on CNN and LSTM to predict whether a hunk in the code change will be accepted by the reviewers.
\citet{hellendoorn2021towards} use the Transformer architecture to solve this task. Furthermore, they also attempt to capture the relation of different hunks in a pull request by encoding each hunk and computing attention scores across diff hunks to fuse the information. \citet{li2019deepreview} formalize automatic code review as a multi-instance learning task, in which each hunk is an instance and the target is to predict whether a pull request will be accepted.

Other researchers focus on tasks related to review comments. 
% Writing review comments is an important but time-consuming task for reviewers.
To save the time reviewers spent on writing reviews related to common issues such as coding style and documentations, \citet{gupta2018intelligent} propose the LSTM-based model DeepMem. DeepMem learns the relations between code changes and review comments, and recommends review comments automatically based on existing code reviews and code changes. \citet{siow2020core} also recommend code reviews in a retrieval-based manner. They propose an LSTM-based multi-level embedding attentional model CORE aiming at capturing the semantic information in both source code and reviews. \citet{tufano2021towards} use deep learning techniques to automate a different task in code review. They train a Transformer model to revise the contributors' code to implement the requirements from the review comments. In the previous work, the researchers usually train a small model for a specific task in the code review process. Differently, we propose a large model pre-trained on four general pre-training tasks for code review, producing superior performance across three different code review tasks.

\subsection{Code Review Datasets}
\label{sec:related_dataset}
To advance the research of automating code reviews activities, researchers also make effort to collect and publish code review datasets. \citet{tufano2021towards}
create two datasets with 17k abstracted code changes for predicting method-level code changes. The Code Review Open Platform (CROP) by \citet{paixao2018crop} is a dataset based on specific Gerrit projects consisting of 507k comments. The dataset from \citet{mukadam2013gerrit} has fewer comments (106k) and provides only metadata of source code. \citet{yang2016mining} propose the dataset with the largest number of comments. But they also provide only metadata of source code. Compared with them,
we propose the largest, multilingual code review dataset with completed information collected from over 1k GitHub repositories with more than 7.9M pull requests in total.

% reduce the workload of reviewers such

\section{Threats to Validity}

\paragraph{Threats to internal validity} relate to the roles played by the model architecture and hyper-parameters setting. We do a small-range grid search on learning rate and batch size settings, leaving other hyper-parameters the same as those in CodeT5 \cite{wang2021codet5}. It is expected that more hyper-parameter tuning would bring more improvements.

\paragraph{Threats to external validity} are mostly related to the dataset we collect and use in this paper. Since the dataset is collected from GitHub, it is built upon only open-source projects, not industrial projects. Besides, code review is often not done by a single reviewer but multiple reviewers, and they could give different comments from different perspectives. But we only collect a single comment for a code change, leading a bias in our dataset.
% even though we explore many ways to do filtering, in the high-quality projects, there still exists review comments that have little to do with the code change, leading noises in our dataset. 

\paragraph{Threats to construct validity} include the rationality of evaluation metrics. We argue that the BLEU score which is widely used in text generation tasks is not suitable for the review generation task. So we conduct a human annotation to better evaluate the results. 
% Based on the feedback of programmers involved in this process, we 

\section{Conclusion}
In this paper, we focus on pre-training techniques for automating code review activities.
We start by formulating three tasks in the code review process with the code change format. 
We collect and organize a large-scale dataset from GitHub for code review pre-training and a benchmark for evaluation on the three tasks. It is the largest dataset in code review scenario, covering nine of the most popular programming languages in GitHub.
Based on this, we introduce CodeReviewer, a transformer-based encoder-decoder model pre-trained on our dataset with four designed pre-training tasks for code review. 
The experimental results demonstrate that our model outperforms the state-of-the-art models pre-trained with source code in all three tasks.

\bibliographystyle{ACM-Reference-Format}
\bibliography{refs}

\end{document}